# Low-profile Button Sensor Antenna Design for Wireless Medical Body Area Networks

Shahid M Ali [A], Cheab Sovuthy [A], Sima Noghanian [B], Qammer H. Abbasi [c], Tatjana Asenova [D], Peter Derleth [D], Alex Casson [E], Tughrul Arslan [F], and Amir Hussain [G]

*Abstract*— a button sensor antenna for wireless medical body area networks (WMBAN) is presented, which works through the IEEE 802.11b/g/n standard. Due to strong interaction between the sensor antenna and the body, an innovative robust system is designed with a small footprint that can serve on- and off-body healthcare applications. The measured and simulated results are in good agreement. The design offers a wide range of omnidirectional radiation patterns in free space, with a reflection coefficient ($S_{11}$) of $-29.30\,(-30.97)$ dB in the lower (upper) bands. $S_{11}$ reaches up to $-23.07\,(-27.07)$ dB and $-30.76\,(-31.12)$ dB, respectively, on the human body chest and arm. The Specific Absorption Rate (SAR) values are below the regulatory limitations for both 1-gram (1.6 W/Kg) and 10-gram tissues (2.0 W/Kg). Experimental tests of the read range validate the results of a maximum coverage range of 40 meters.

*Clinical Relevance*— Wireless Body Area Network (WBAN) technology allows for continuous monitoring and analysis of patient health data to improve the quality of healthcare services.

## I. Introduction

Wireless medical body area networks (WBAN) have received high interest in recent years. One of the most significant components of WMBAN is the wearable antenna, which is in high demand for applications such as health monitoring, physical training, navigation, as shown in Figure 1. In WMBAN, most of the research focuses on the antenna design and its implementation [1]. Some studies look at the materials' properties and stability, as well as how they affect the antenna performance. Other research examines the impact of the human body on antenna performance and SAR values. Wearable antennas are designed to be flexible. In many of these designs, the antenna is made of textile materials and takes planar shapes. However, fully textile antennas have some shortcomings. Their performance can be drastically affected by bending, stretching, crumpling and other deformations. In contrast, small rigid and semi-rigid designs such as button antennas that can be worn on shirts and be a part of jeans offer rigidity to maintain antenna performance. A button sensor antenna using copper conductive materials can increase the antenna's efficiency [2-3]. Numerous studies on button antennas have been published in the last few years [2-9]. Thanks to the fast designs of low-power-integrated circuits, single-chip on-body sensor antennas can be used for long time.

This paper proposes a new asymmetrical dual-band design that can fit a button shape and be used as a sensor antenna or a terminal sensor antenna working as part of a single chip sensor antenna. The antenna offers a small footprint of 45 mm × 45 mm. The antenna exhibits a pattern like a monopole antenna at 2.45 GHz, while in the 5.6 GHz, it exhibits a broadside pattern. Therefore, both on- and off-body healthcare applications can be served by this antenna.

This proposal type is organized as follows. Section II briefly describes the design requirement and related studies. In Section III, the button sensor antenna's performance is discussed. Finally, Section IV presents some concluding remarks

[1]* Research supported by the UK EPSRC COG-MHEAR programmed grant (Grant no. EP/T021063/1)

A. Shahid M Ali and Cheab Sovuthy are with Department of Electrical and Electronic Engineering, Universiti Teknologi, PETRONAS, Malaysia; shahid_17006402@utp.edu.my; sovuthy.cheab@utp.edu.my

B. Sima Noghanian is with CommScope; sima_noghanian@ieee.org

C. Qammer H. Abbasi is with the James Watt School of Engineering, University of Glasgow, UK; Qammer.Abbasi@glasgow.ac.uk

D. Tatjana Asenova and Peter Derleth are with Sonova AG, Switzerland; tatjana.asenov@sonova.com , peter.derleth@sonova.com

E. Alex Casson is with the University of Manchester; UK, alex.casson@manchester.ac.uk

F. Tughrul Arslan is with the University of Edinburgh, UK; T.Arslan@ed.ac.uk

G. Amir Hussain is with the School of Computing, Edinburgh Napier University, UK; A.Hussain@napier.ac.uk

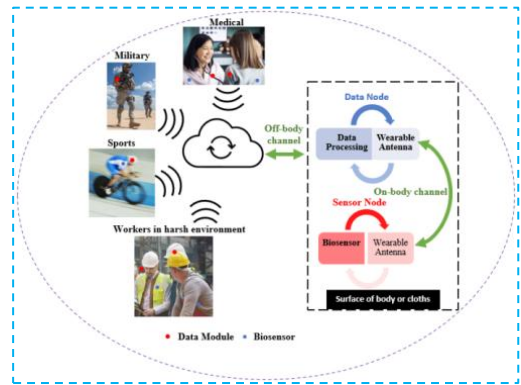

Figure 1. Healthcare design in WMBAN system.

## I. System Requirement

This section describes the system requirements and design choices. The button antenna and backbone electronics are both investigated. Further, an appropriate design process is chosen, as shown in Fig. 2.

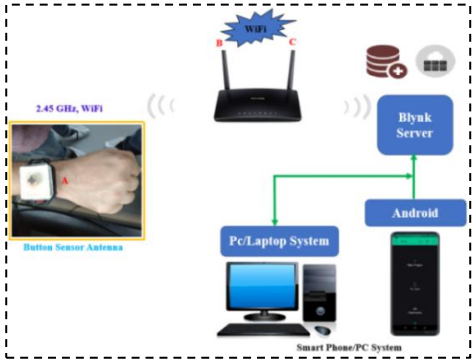

Fig. 2. System architecture for the button sensor antenna.

## A. Button Antenna Topology

A button sensor antenna as shown in Fig. 3, using the CST Microwave Studio (MWS). On the PCB button Rogers's substrate (RT/Duroid 5880); the radio-frequency (RF) antenna was constructed. This substrate has a thickness of 1.574 mm, a relative permittivity ($\varepsilon_r$) of 2.2, and a loss tangent (tan δ) of 0.0009 S/m, which was placed on a 1.50 mm felt substrate with a $\varepsilon_r$ of 1.4 and a tan δ of 0.044 S/m. There is a 3.76 mm air gap between the PCB layer and the felt substrate. The conductive ground layer is made of ShieldIt, with an area of 45 mm × 45 mm, a thickness of 0.17 mm, with conductivity of $1.18 \times 10^5$ S/m. ShieldIt layer was glued on the bottom side of the felt substrate.

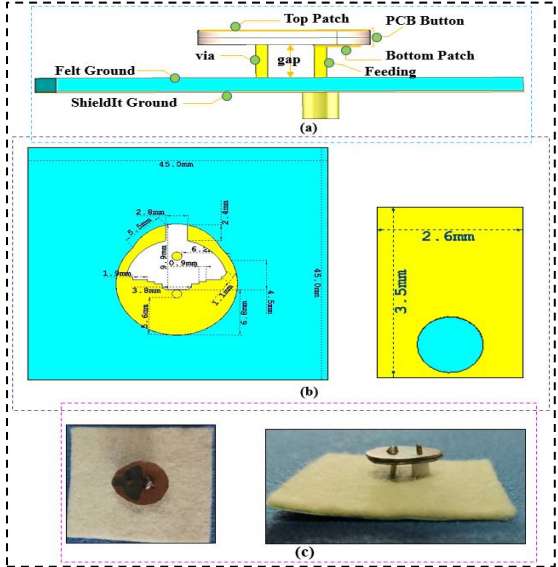

Fig. 3. Structure and dimensions of the BSA (dimensions in the given mm): (a) the side view; (b) the top view; and (c) the fabricated BSA.

The conductive parts on the top side of the PCB with a radius of 8 mm are as follows: a pin-fed patch on the bottom side of the PCB, and an asymmetrical capacitive patch connected to the bottom patch on the top side, which is also short-circuited to the ground plane by a shorting via. The coaxial feed and the shorting via have diameters of 1.27 mm and 1.22 mm, respectively. Glue was used to adhere the ShieldIt ground plane to the felt substrate. For the SMA connector and its galvanic connection to ShieldIt layer, a through-hole was made. In addition, the center asymmetrical slot was capacitively coupled to the patch and added extra load. A staircases shape was utilized to achieve a compact size [30]. The radiating patch located on the top side of the PCB can be considered as a circular loop. Its radius ($a$) can be calculated by:

$$a = \frac{F}{\left\{1+\frac{2h}{\pi\varepsilon_r F}\left[ln ln\left(\frac{\pi F}{2h}\right)+1.7726\right]\right\}^{1/2}} \quad (1)$$

$$F = \frac{8.791 \times 10^9}{f\sqrt{\varepsilon_r}} \quad (2)$$

Where $h$ is the thickness of the substrate and $\varepsilon_r$ is its dielectric constant. Equation (1) does not take the fringing effect into consideration. Since fringing makes the patch electrically larger, the effective radius ($a_e$) of the patch must be used which is given by [3]:

$$a_e = a\left\{1+\frac{2h}{\pi\varepsilon_r a}\left[ln\,ln\left(\frac{\pi a}{2h}\right)+1.7726\right]\right\}^{1/2} \quad (3)$$

$$f_r = \frac{1.8412c}{2\pi a_e \sqrt{\varepsilon_r}} \quad (4)$$

Where $c$ the free-space is speed of light, and $f_r$ is the resonance frequency.

## B. Design of Printed Circuit Board

Eagle, version 7.7.0 design tool was used to create the schematic design and a printed circuit board. After validating the simulation and breadboard prototype, the schematic design was transferred to a PCB prototype, as shown in Fig. 4.

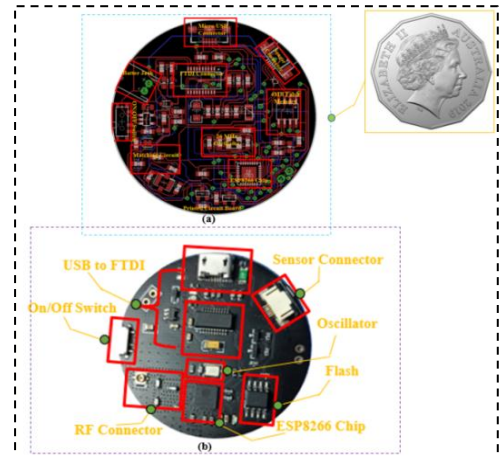

Fig. 4. PCB Design. (a) Simulated PCB design, and (b) fabricated PCB prototype.

To keep the schematic design as small as possible, surface mount components are used. The diameter of this PCB is 45 mm. The prototype was optimized and designed on FR4 material, which has a dielectric constant of 4.2, a loss tangent of 0.02, and a thickness of 1.6 mm. The developed wireless sensor module was then properly programmed for RSSI at the ISM band at 2.45 GHz using the IEEE 802 b/g/n standard and incorporated with a button antenna via a T-shaped matching circuit. The ESP8266 is a Wi-Fi-enabled integrated circuit. The circuit-based modules are widely used to control devices via the Internet. The 26 MHz Crystal Oscillator, 4MB Flash memory, USB to FTDI connector, Step-down regulator (3.3 V), power supply, and 2.4 GHz RF transceiver are all included in ESP8266 EX wireless module. An enclosure as shown in Fig. 5, similar to a hand watch cover, was designed and 3D printed. The button antenna was placed on top of the cover module, and they were connected through U.FL connector.

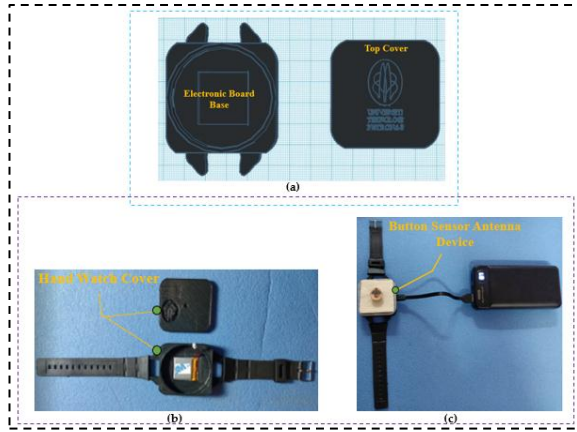

Fig. 5.3D covers of a wearable button sensor antenna. (a) Simulated cover, and (b) fabricated cover.

## II. RESULTS AND DISCUSSION

To find the optimum design parameters a parametric study was conducted while yielding a realizable structure. The time-domain solver in CST MWS was used for the design and parametric tolerance analysis. Simulations and measurements were done in the frequency 1 – 7 GHz.

### A. Button Sensor Antenna Locations

Figure 6 depicts the simulated reflection coefficients ($S_{11}$) of the button sensor antenna, as well as the impacts of feeding locations on the $S_{11}$.

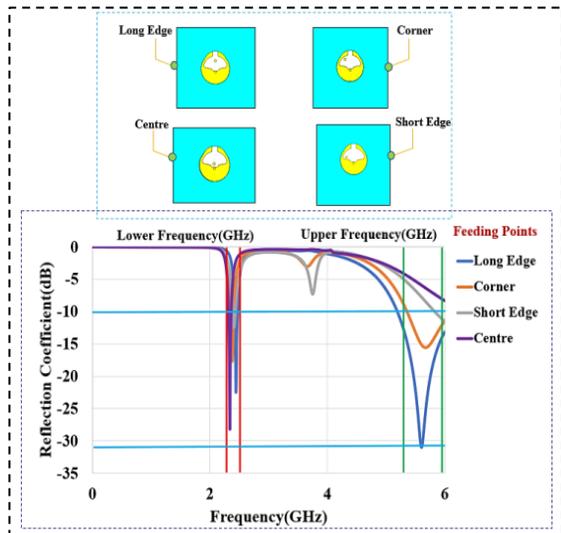

Fig. 6. $S_{11}$ of the BSA for different feed locations.

It can be seen how changing the button sensor antenna's feeding point has a considerable impact on the impedance matching level and resonant frequency bands. Due to the presence of the lossy tissues, the radiation pattern has become wider with high stability for the on-body case. As illustrated in Figure 7, an omnidirectional radiation pattern is obtained.

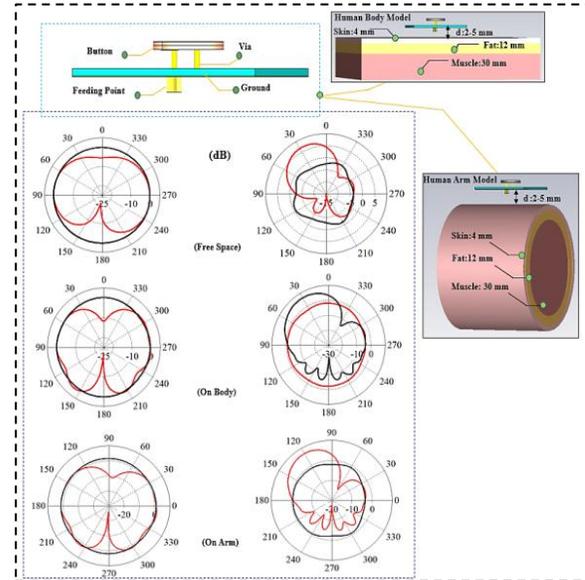

Fig.7.Radiation patterns of different devices BSA at lower (left) and upper band (right) in various scenarios. Solid black line: H Plane, red line: E Plane.

### B. Body Phantom Models

To deploy wearable designs, it is necessary to study their performance while being placed on the body and tissues. For examples, some of the models of interest are the chest and the arm. The performance of the design on the models of the human chest and arm (Fig. 8) will be presented in this section. For the chest model, a flat 3D-layered body model, consisting of skin, fat, and muscle layers, was used. The chosen dimensions of this model are $200 \times 200 \times 50$ mm$^3$, and the thicknesses of the layers are as follows (in mm): skin: 4, fat: 8, and muscle: 30. The muscle layer has $\varepsilon_r$ of 52.7 and 48.2, and conductivity of 1.95 and 6.0 S/m, at 2.4 GHz and 5.6 GHz, respectively.

For modeling of the button sensor antenna placement on the arm, a simple layered model was utilized. For the on-body communication described in [4], a phantom with a radius of 50 mm and a length of 150 mm was chosen for the human arm. Although the antenna's center frequency is affected by the presence of the lossy tissue, it still operates in the appropriate frequency band. The various possible bending conditions that we considered are depicted in Fig. 9. Therefore, we can conclude that the device has robustness against a bending radius of R = 50 mm. Fig. 10 depicts $S_{11}$ for the antenna in free space and on body phantoms. Simulated and measurement results are in good agreement. Furthermore, it is very important to examine the link budget of the wearable devices in body-centric communications to ensure a reliable link budget with the range of 40 meters, as shown in Figure 25.The SAR average values are calculated on every 10 g and 1 g of tissues [4], and they show that the device can be safely operated near the human body tissues at 5 mm.

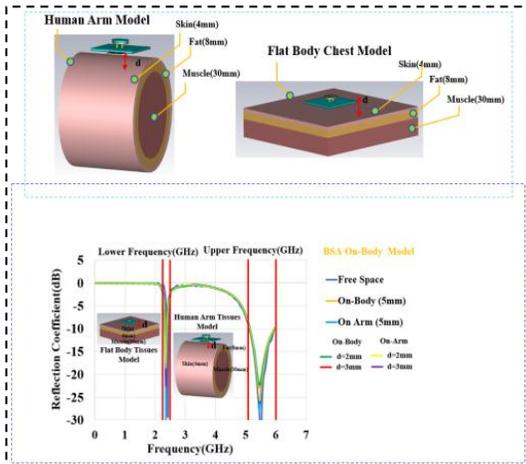

Fig. 8.$S_{11}$ in free-space, on-chest flat model, and arm model (in mm, 2, 3, and 5).

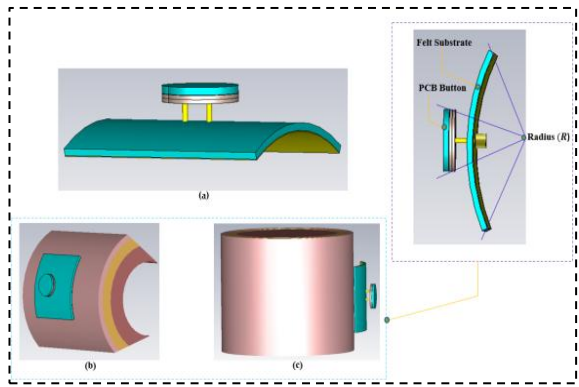

Fig. 9. Effect of bending of the antenna,(a) free space, ( b) layered chest phantom (size, 200 ×200 × 30 mm$^3$), and (c) arm cylindrical phantom (50 mm diameter).

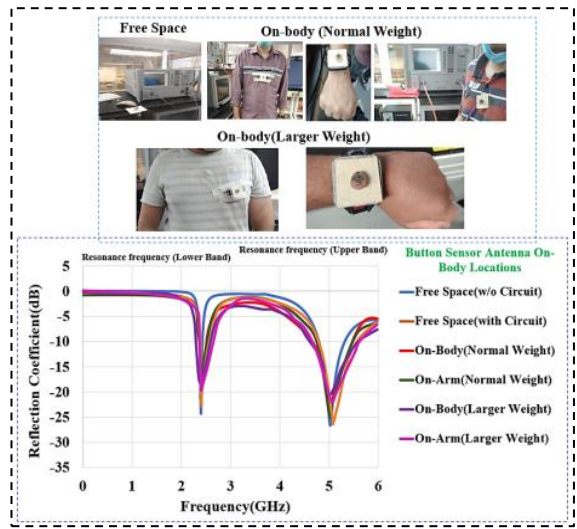

Fig.10.$S_{11}$ in free-space and on-body.

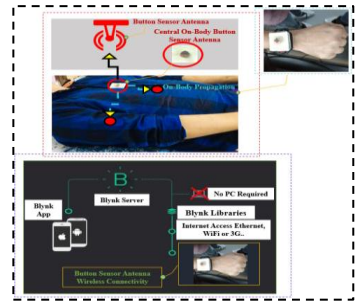

Fig. 11. Measurement scenarios for on-body communication.

### III. CONCLUSION

In this paper, an innovative button shaped sensor antenna for healthcare WMBAN applications was presented, which operates at two frequency bands of 2.45 GHz and 5.6 GHz. The BSA is small, with a button form and an area of 45 ×45 mm$^2$. This is integrated into a circular-shaped wireless sensor module with a 45 mm diameter that can be encased in a 3D cover like a watch, or embedded in clothing. The proposed design demonstrates high stability in terms of operation resonance frequency, impedance matching level, radiation patterns, less attenuation through materials, and reduced SAR values. The simulation and measurements results are found to match well. The antenna can be used in a range of off-body and on-body healthcare applications. Ongoing work is evaluating its use in a real-time hearing-assistive technology prototype [10].